\documentstyle [11pt]{article}

\font\tenrm=cmr10
\font\tenit=cmti10
\font\elevenbf=cmbx10 scaled\magstep 1
\font\elevenrm=cmr10 scaled\magstep 1
\font\elevenit=cmti10 scaled\magstep 1

\newcommand{\ra}{\rightarrow}
\newcommand{\Ra}{\Rightarrow}
\newcommand{\be}{\begin{equation}}
\newcommand{\ee}{\end{equation}}
\newcommand{\ba}{\begin{array}}
\newcommand{\ea}{\end{array}}
\newcommand{\bc}{\begin{center}}
\newcommand{\ec}{\end{center}}
\newcommand{\bea}{\begin{eqnarray}}
\newcommand{\eea}{\end{eqnarray}}
\renewenvironment{thebibliography}[1]
 { \elevenrm
   \begin{list}{\arabic{enumi}.}
    {\usecounter{enumi} \setlength{\parsep}{0pt}
     \setlength{\itemsep}{3pt} \settowidth{\labelwidth}{#1.}
     \sloppy
    }}{\end{list}}

\begin{document}
\begin{picture}(5,2.5)(-315,-90)
\put(12,-90){UPR--0574T}
\put(17,-105){June 1993}
\end{picture}
\begin{center}
\vglue 0.6cm
{\elevenbf        \vglue 10pt
               BREAKDOWN OF DUALITY IN (0,2) SUPERSTRING
               MODELS\footnote{Talk presented at the International Workshop:
      {\sl Recent Advances in the Superworld},
      Houston Advanced Research Center, The Woodlands, Texas, April 1993.}\\}
\vglue 1.0cm
{\tenrm JENS ERLER\footnote{e-mail: erler@langacker.hep.upenn.edu} \\}
\baselineskip=13pt
{\tenit Department of Physics, David Rittenhouse Laboratory,
        University of Pennsylvania, \\}
\baselineskip=12pt
{\tenit 209 South 33rd Street, Philadelphia, PA 19104, U.S.A.\\}
\vglue 0.8cm
{\tenrm ABSTRACT}
\vglue 0.3cm
\end{center}
{\rightskip=3pc
 \leftskip=3pc
 \tenrm\baselineskip=12pt
After pointing out the role of the compactification lattice for
spectrum calculations in orbifold models, I discuss
modular discrete symmetry groups for $Z_N$ or\-bi\-folds.
I consider the $Z_7$ orbifold as a nontrivial example of
a (2,2) model and give the generators of the modular group for
this case, which does not contain $[SL(2,{\bf Z})]^3$ as had been
speculated. I also discuss how to treat cases where quantized Wilson
lines are present. I consider in detail
an example, demonstrating that quantized Wilson lines affect the modular
group in a nontrivial manner. In particular, I show that it is possible for
a Wilson line to break $SL(2,{\bf Z})$.}
\vglue 0.6cm
{\elevenbf\noindent 1. Introduction}
\vglue 0.1cm
\baselineskip=14pt
\elevenrm
The space of string vacua is locally parametrized by moduli,
which are marginal deformations of the underlying conformal field
theory (CFT).
In the low energy effective theory the moduli correspond to vacuum expectation
values of massless scalar fields that have flat potentials to all
orders in perturbation theory.
An intriguing feature of string compactifications is that the
natural parametrization of the moduli space label the points in a
redundant way, since all physical quantities are invariant under the action
of some discrete group acting on the moduli.

The motivations for studying this modular discrete symmetry groups for
compactifications of the heterotic string are manifold. The most obvious
one is that its knowledge allows a restriction to a fundamental domain of
moduli space, redu\-cing the number of string vacua. Next, being an exact
symmetry it should persist in any approximation, and in particular in
the low energy effective supergravity theory. This constrains the
superpotential, the gauge kinetic functions, etc. Assuming the exactness
of this symmetry even after taking into account nonpertubative string effects,
gives a powerful tool for the construction of nonpertubative
potentials$^1$. They can fix the moduli vacuum expectation values (vev's)
and lift the vacuum degeneracy.
Recent discussions of soft supersymmetry breaking terms and minimal string
models were restricted to the modular group $SL(2,{\bf Z})$. It will be one
of the main results of this talk that this group plays a much smaller
role as a symmetry in moduli space as previously believed.
Finally, modular symmetries may also play an important role
in the context of gauge coupling unification. Here threshold corrections
potentially explain the discrepancy between the reduced Planck scale and
the unification scale. Again, the relevant threshold terms have to transform
appropriately under the given symmetry group$^2$.

I will discuss these modular symmetries for orbifold models.
They represent a simple construction and the orbifold CFT is exactly solvable.
This implies that the discrete symmetry groups are exactly derivable, as well,
and that the phenomenology of orbifold models can be pushed forward
very far. At the same time ``quasi realistic'' models with
3 fermion generations transforming under
$SU(3) \times SU(2) \times U(1)^n$ belong to this class$^3$.
A natural hierarchy of fermion masses arises due to the fixed point
structure of the twisted sectors. Finally, exact deformations of fermionically
constructed models can be obtained via the $Z_2 \times Z_2$ orbifold.
\vglue 0.5cm
{\elevenbf\noindent 2. Symmetric $Z_N$ Orbifolds}
\vglue 0.4cm
The possible point groups leading to $D=4$, $N=1$ supersymmetric
$Z_N$ or\-bi\-folds are well known$^4$:
\be
Z_3,\quad Z_4,\quad Z_6,\quad Z_6^{\prime},\quad Z_7,\quad Z_8,\quad
 Z_8^{\prime},\quad Z_{12},\quad Z_{12}^{\prime}.
\ee
However, in order to find all the corresponding {\elevenit models\/} a
classification
of the possible compactification lattices is necessary. As an example of how
the underlying lattice affects basic properties of the model consider the
$Z_4$ orbifold. It can be realized e.g.\ using three copies of the
root lattice of $SO(4)$ with a twist simultaneously acting as
$90^o$ rotations in two copies and as a reflection in the last one:
$$\Lambda_1 = [SO(4)]^3:\quad\quad
e_{1,3} \ra e_{2,4} \quad e_{2,4} \ra -e_{1,3} \quad e_{5,6} \ra -e_{5,6}.$$
Alternatively, one can utilize two copies of the root lattice of
$SU(4)$ with Coxeter twists acting in them:
\be
\Lambda_2 = [SU(4)]^2:\quad\quad
e_{1,4} \ra e_{2,5} \quad e_{2,5} \ra e_{3,6} \quad
e_{3,6} \ra -\sum_{i=1}^3 e_{i,i+3}.
\ee
It has been shown that for two models to be equivalent,
there must be a matrix $M \in GL(d,{\bf Z})$ with respect to which the
integer valued twist (lattice basis) matrices $\theta$ are similar$^5$, i.e.
\be
\exists \quad M \in GL(d,{\bf Z}) \quad {\rm with} \quad
            \theta^{\prime} = M \theta M^{-1}.
\ee
For the case at hand such a matrix cannot be found. What does that mean for the
spectra?

Clearly, in the untwisted sector the lattice $\Lambda$ (winding modes) and
its dual $\Lambda^*$ (momenta) only affect the massive string modes.
The untwisted massless states are the same.
Since there are no fixed tori in the first twisted sector, it is
completely independent of $\Lambda$. The number of generations coming
from this sector equals the number of fixed points, which in turn can be
determined with help of the Lefschetz fixed point theorem. In our case
we find
\be
{\rm Det} (1-\theta)=16.
\ee
The second twisted sector, however, {\elevenit does\/} depend on $\Lambda$ due
to the complex plane which is left fixed under the twist action. The number
of generations coming from such a sector is given by the number of
fixed tori.

In the case of the
$[SO(4)]^3$ lattice the fixed plane is orthogonal to the twisted directions.
This implies that the Lefschetz fixed point theorem is again applicable
if restricted to the twisted directions. One finds 16 fixed tori. A closer
investigation of twist phases reveals that 10 of them correspond to
generations and 6 to antigenerations.
In contrast, in the $[SU(4)]^2$ model the Lefschetz theorem cannot be
employed. Explicit construction of the fixed tori gives rise to four
generations and no antigenerations.

The same result can be obtained by computing the one loop partition function.
The relevant projections in the two cases are$^6$
\bea
\Lambda_1: {1\over 4} [16 + 4 \Delta + 16 \Delta^2 + 4 \Delta^3], \label{d1} \\
\Lambda_2: {1\over 4} [ 4 + 4 \Delta +  4 \Delta^2 + 4 \Delta^3], \label{d2}
\eea
with $\Delta = + 1$ for generations and $\Delta = - 1$ for antigenerations.
The degeneracy factors in Eqs.~(\ref{d1}) and (\ref{d2}) are
different because the twist acts as a rotation in the even, selfdual
lattice composed of winding and momentum states: In the orthogonal
case of $\Lambda_1$ the
invariant sublattice is even and selfdual itself and thus the volume factor
which one encounters when performing modular world sheet transformations is
trivially one. On the other hand, a nontrivial volume factor is found for
$\Lambda_2$.

{}From Eqs.~(\ref{d1}) and (\ref{d2}) it is also obvious, that the Euler number
determining the number of net generations is independent of the underlying
compactification lattice.
This one indeed expects, since this number can be
computed by a formula conjectured by Dixon, Harvey, Vafa and Witten$^4$
and proved by Markushevich, Olshanetskii and Perelomov$^7$, which only
uses twist eigenvalues.

Because of their appealing phenomenology and a considerable confusion in the
literature, it was worthwhile to construct all supersymmetric, symmetric
$Z_N$ orbifolds with (2,2) world sheet supersymmetry$^6$. The resulting
18 models have different massive and {\elevenit massless\/} spectra, different
modular symmetry groups and consequently different phenomenologies.
\vglue 0.5cm
{\elevenbf \noindent 3. Duality and Discrete Symmetry Groups}
\vglue 0.2cm
{\elevenit\noindent 3.1. Overview}
\vglue 0.1cm
To begin the discussion of discrete symmetries in moduli space, consider as a
simple example the $Z_3$ orbifold in two dimensions. The twist acts in the root
lattice of $SU(3)$ and the scaling deformations thereof. There is also a
continuous antisymmetric tensor field compatible with the twist:
\be
g=R^2 \left( \ba{cc} 2 & -1 \\ -1 & 2 \ea \right), \quad\quad
b_{ij}=    \left( \ba{cc} 0 & -b \\  b & 0 \ea \right).
\ee
The orbifold radius $R$ and the antisymmetric tensor parameter $b$,
which are the real untwisted moduli of the model, can be
combined to the complex parameter
\be
\lambda = b + i\sqrt{\rm Det\,g},
\ee
which takes values in the complex upper half plane.

Duality symmetry is the statement that the exchange of winding and momentum
quantum numbers,
\be \label{dual1}
n \ra W m, \quad\quad m \ra {W^{-1}}^T n,
\ee
and a simultaneous transformation of the background,
\be \label{dual2}
g \pm b \ra {W^{-1}}^T {1\over g \pm b} W^{-1},
\ee
lead to the same physical theory. Here the matrix $W \in GL(2,{\bf Z})$
has to satisfy the condition$^8$
\be \label{WMAT}
   \theta W = W {{\theta}^{-1}}^T.
\ee
The transformation of the complex modulus
$\lambda$ takes the particularly simple form of an inversion,
\be
\lambda \ra -1/\lambda .
\ee
Inclusion of the {\elevenit axionic\/} shift symmetry,
\be
b \ra b + 1 \quad \Ra \quad \lambda \ra \lambda + 1,
\ee
completes the holomorphic symmetry group$^9$ $PSL(2,{\bf Z})$.
As shown by Lauer, Mas and Nilles$^{10}$ these symmetries are also respected
by the correlation functions. Finally, there is a nonholomorphic
generator$^{11}$,
\be \label{nonhom}
b \ra -b \quad \Ra \quad \lambda \ra -\bar{\lambda},
\ee
which induces an exchange of particles and antiparticles in the twisted
sectors.

The results of this example can be summarized in compact form by noting
that the global structure of the moduli space is given by
\be
{\cal M}_{T_2/Z_3} = {{SU(1,1)\over U(1)}\over SU(1,1;{\bf Z})\times Z_2}.
\ee
Written this way, for the holomorphic part one can state the result in
saying that the discrete symmetry group is given by the maximal discrete
subgroup of the group characterizing the local structure of the moduli space.

The question arises whether this is a general feature. Consider the local
structures of the moduli spaces of the $Z_N$ orbifolds$^{12}$:
\be
{\cal M}_{Z_3} = {SU(3,3)\over SU(3)\times SU(3)\times U(1)}
\ee
\be
{\cal M}_{Z_{4,6}}   = {SU(2,2)\over SU(2)\times SU(2)\times U(1)}
                    \times \left[ {SU(1,1)\over U(1)} \right]^n\quad n=1,2
\ee
\be
{\cal M}_{Z_{N \geq 7}} = \left[ {SU(1,1)\over U(1)} \right]^n\quad n=3,4.
\ee
It has often been assumed that, in fact, the maximal discrete subgroups
of the above local structures describe the modular symmetries. Also,
extended versions for cases including Wilson lines leading to (0,2) models with
broken $E_6 \times E_8$ gauge symmetry are used. For instance,
for the $Z_3$ orbifold one would expect the group $SU(3,3;{\bf Z})$, and,
similarly, $[SU(1,1;{\bf Z})]^3 \cong [PSL(2,{\bf Z})]^3$ for $Z_7$.
Actually, this hypothesis has
been confirmed for the $Z_3$ case without Wilson lines$^{13}$. Let us now
turn to the more complicated case of the $Z_7$ orbifold.
\vglue 0.2cm
{\elevenit\noindent 3.2. $Z_7$ Orbifold}
\vglue 0.1cm
It is defined as the Coxeter twist
\be
e_i \ra e_{i+1} \quad\quad i=1,\dots,5\quad\quad\quad e_6 \ra -\sum_{i=1}^6 e_i
\ee
in the $SU(7)$ root lattice and the three metrical deformations thereof.
Analogously, three independent parameters of the antisymmetric tensor
background are compatible with this twist. Again drastic simplifications
occur when using complex moduli, e.g.
\be
t_1 = - i\tan({\pi \over 7}) [g_1 +{s_1^2\over s_3^2}g_2 +{s_2^2\over
s_3^2}g_3]
    + b_1 + {s_2 \over s_1} b_2 + {s_3 \over s_1 } b_3, \nonumber
\ee
\be
t_2 = - i\tan({2\pi\over 7}) [g_1 +{s_2^2\over s_1^2}g_2 +{s_3^2\over
s_1^2}g_3]
    + b_1 - {s_3 \over s_2 } b_2 - {s_1 \over s_2 } b_3, \nonumber
\ee
\be
t_3 = - i\tan({3\pi\over 7}) [g_1 +{s_3^2\over s_2^2}g_2 +{s_1^2\over
s_2^2}g_3]
    + b_1 - {s_1 \over s_3 } b_2 + {s_2 \over s_3 } b_3,
\ee
with the abbreviations $s_k := \sin{2\pi k \over 7}$ and
$c_k := \cos{2\pi k \over 7}$. These moduli are normalized in such a way
that one of the axionic shift symmetries ($b_1 \ra b_1 + 1$) takes a
simple form,
\be
T: (t_1,t_2,t_3) \ra (t_1+1,t_2+1,t_3+1).
\ee
Similarly, a matrix $W$ (cf. Eqs.~(\ref{dual1}) and (\ref{dual2})) can be
found$^{14}$ such that a $PSL(2,{\bf Z})$ structure arises,
\be
S: (t_1,t_2,t_3) \ra (-1/t_1,-1/t_2,-1/t_3).
\ee
However, no transformation changing only one {\elevenit individual\/} complex
modulus can be found, no matter which definition for the $t_i$ one
uses$^{14}$. This already reveals that there is no simple $[SL(2,{\bf Z})]^3$
symmetry which one would expect on the basis of the aforementioned conjectures.
In contrast, a rich structure arises. It is best described by introducing
another type of symmetry generators:

Transforming quantum numbers and background according to
\be \label{dual3}
n \ra V n, \quad\quad m \ra {V^{-1}}^T m,
\ee
and
\be \label{dual4}
g \pm b \rightarrow \ra {V^{-1}}^T (g \pm b) V^{-1},
\ee
respectively, gives rise to a symmetry, presupposing the matrix
$V \in GL(2,{\bf Z})$ satisfies
\be \label{VMAT}
   \theta V = V {\theta}^p.
\ee
$p$ is an integer which is in general allowed to be greater than one$^{15}$.
However, $p$ is required$^{11}$ to have no common divisor with the
twist order $N$. The sig\-ni\-fi\-cance of transformations with $p > 1$ is
that they correspond to nonholomorphic transformations
(cf. Eq.~(\ref{nonhom})).

For the $Z_7$ case we found the transformation
\be
R: (t_1,t_2,t_3) \ra (-2 c_1 \bar{t_2},-2 c_2 t_3,-2 c_3 t_1).
\ee
$R^3$ yields just complex conjugation.
It is interesting that the inclusion of nonholomorphic transformations
actually reduces the number of symmetry {\elevenit generators\/} since it
relates holomorphic ones. For instance the shift symmetries $T_{2(3)}:=
b_{2(3)} \ra b_{2(3)} + 1$ can be generated like ($T=T_1$),
\bea
R T_1 R^{-1} =: T_3, \\
R T_3 R^{-1} =: T_2.
\eea
Similarly ($S=S_1$),
\bea
S_2 := R^2 S_1 R^{-2}, \\
S_3 := R S_1 R^{-1}.
\eea
This way more $PSL(2,{\bf Z})$ subgroups arise, and in fact, there is an
infinity of them. However, no pair of these subgroups mutually commute,
again showing that there is no $[SL(2,{\bf Z})]^3$.

The three generators $R$, $S$ and $T$ are sufficient to generate the whole
symmetry group. There are many relations between them, e.g.\
\be \ba{c}
   S^2 = (ST)^3 = R^6 = {\bf 1} \\ \\
   (TR)^6 = (TR^3)^2 = (SR^3)^2 = {\bf 1} \\ \\
   (SRTR^{-1})^7 = {\bf 1} \\ \\
   etc.
\ea \ee
It is interesting to note that whereas duality is known to relate small
and large radii, the combination $SR^2SR^{-2}$ yields a
{\elevenit rescaling symmetry}:
\be
(t_1,t_2,t_3)\ra
({1\over 4c_1^2}\, t_1,{1\over 4c_2^2}\, t_2,{1\over 4c_3^2}\, t_3).
\ee
It is of infinite order, transforming two of the moduli to
smaller and one of them to larger values.

All holomorphic transformations are of the form
\be \label{tra}
t_k \ra {a_k + b_k t_k\over c_k + d_k t_k}.
\ee
The 12 integers appearing in Eq.~(\ref{tra})
satisfy three nonlinear relations, generalizing the
determinant condition $ad-bc=1$ in the case of $SL(2,{\bf Z})$. They enter
the matrix representation of the symmetry group at hand$^{14}$.

We checked that the symmetries are respected by the correlation functions,
which are now available with their complete moduli
dependance$^{10,16}$.
\vglue 0.5cm
{\elevenbf \noindent 4. Discrete Background Fields and (0,2) Models}
\vglue 0.2cm
{\elevenit\noindent 4.1. Orbicircle $\bigotimes$ Circle }
\vglue 0.1cm
In this section I describe how discrete background fields break
the mo\-du\-lar symmetry group down to a subgroup.
The simplest example is the product theory of an orbicircle (a circle
with its $Z_2$ symmetry divided out) with a circle. The twist matrix is
given by
\be
\theta = \left( \ba{cc} 1 & 0 \\ 0 & -1 \ea \right),
\ee
giving rise to the metric
\be
g = \left( \ba{cc} R_C^2 & 0 \\ 0 & R_O^2 \ea \right),
\ee
characterized by the orbifold radius $R_O$ and the circle radius $R_C$.
In this example the antisymmetric tensor field does not possess a
continuous parameter and, in contrast to the two-dimensional
$Z_3$ orbifold discussed earlier, does {\elevenit not} correspond to a modulus.
It only takes discrete values,
\be
b_{ij} = \left(\ba{cc} 0 & k \\ -k & 0 \ea \right), \quad\quad
     k \in {1\over 2} {\bf Z}.
\ee

Two different cases have to be distinguished. In the first, one has
$k \in {\bf Z} \cong 0$. This corresponds to a proper direct product
theory of the orbicircle and the circle. Thus it is clear that the
symmetry transformations consist of the two independent duality involutions
\be
R_C \ra 1/R_C \quad\quad {\rm\bf and/or}  \quad\quad R_O \ra 1/R_O,
\ee
giving rise to $Z_2 \times Z_2$.

The second case has $k \in {\bf Z} + 1/2$. The nontrivial antisymmetric
tensor field has two effects. Its presence only allows for a
{\elevenit simultaneous\/} inversion of the radii,
\be
R_C \ra 1/2 R_C \quad\quad {\rm\bf and}  \quad\quad R_O \ra 1/2 R_O,
\ee
and changes the self-dual point. In other words, it breaks$^{5}$ the
$Z_2 \times Z_2$ symmetry down to $Z_2$.
\vglue 0.2cm
{\elevenit\noindent 4.2. $Z_3$ Orbifold $\bigotimes$ Gauge Lattice}
\vglue 0.1cm
As an example with more phenomenological applicability consider the
$Z_3$ orbifold with a {\elevenit quantized Wilson line\/} turned on.
In complete analogy to the toy example in the preceding subsection,
we have the combination of an orbifold with a torus connected through
a discrete background field. The presence of quantized Wilson lines is
phenomenologically highly desired. They lift the degeneracy of the twisted
sectors' fixed points, thereby reducing the number of fermion generations
from 36 to a smaller number and three generation models can be
found$^{3}$. At the same time the $E_6 \times E_8$ gauge symmetry
is broken to realistic gauge groups. The lesson to be
learned here is that they also break the discrete symmetry group in
moduli space. Most surprisingly, they break duality symmetry. The $Z_2$
duality generator has to be replaced by an infinite order
generator acting on the background as
\be
A: g+b-\Delta_{1/3} \ra {W^{-1}}^T {1\over 9(g \pm b + \Delta_{1/3})}W^{-1},
\ee
where
\be
\Delta_{1/3} := {1\over 3} \left( \ba{cc} 0 & 1 \\ -1 & 0 \ea \right).
\ee

In order to show that the remaining symmetry is in fact a subgroup
of $SL(2,{\bf Z})$, I introduce the rescaled complex modulus
\be
  \lambda^{\prime} = {1\over 3} \lambda =  {1\over 3} (b + i\sqrt{\rm Det\,g}).
\ee
The $SL(2,{\bf Z})$ transformations w.r.t.\ $\lambda^{\prime}$ are defined as
\bea
   \sigma: \hspace{25pt} \lambda^{\prime} &  \ra -{1\over \lambda^{\prime}}, \\
   \tau:   \hspace{25pt} \lambda^{\prime} &  \ra \lambda^{\prime} + 1.
\eea
Now one can write
\bea
   A := \tau \sigma \tau, \\
   B := \tau^3,
\eea
where $B$ corresponds to the axionic shift symmetry.
The $SL(2,{\bf Z})$ relations $S^2 = (ST)^3 = 1$ have to be replaced by
$(AB)^3 = 1 $ with no order two relation.

Thus {\elevenit quantized Wilson lines break} $SL(2,{\bf Z})$
{\elevenit to a subgroup}$^{11,14}$. On the other hand,
the ``canonical'' duality transformation is {\elevenit not} modular and
leads to an asymmetric orbifold$^{5}$.
\vglue 0.5cm
{\elevenbf \noindent 5. Conclusions and Outlook}
\vglue 0.4cm
It has become clear that only in a very limited number of cases the
modular group $SL(2,{\bf Z})$ is realized as a symmetry in  moduli space.
This has to be compared with the result of an investigation of the mirror
manifold of the quintic threefold$^{17}$. As in our case, there is a shift
symmetry but no order two generator (duality), which in their case is replaced
by an order five generator.
The phenomenological investigations using $SL(2,{\bf Z})$ or the
``maximally discrete symmetry hypothesis'' have to be generalized to
other groups.

The classification of (2,2) orbifolds necessarily involved the
construction of all possible compactification lattices, since they can
affect the massless spectra. The list of symmetric orbifolds is now
complete and the one of asymmetric orbifolds (point groups {\elevenit and\/}
lattices) is under way.
\vglue 0.5cm
{\elevenbf \noindent 5. Acknowledgements \hfil}
\vglue 0.4cm
It is a pleasure to thank D. Jungnickel, Albrecht Klemm,
J. Lauer, J. Mas, M. Spali\~nski, S. Stieberger and especially H. P. Nilles
with whom I collaborated on the subjects presented here. I also thank S.
Theisen for stimulating discussions. I acknowledge support by Evangelisches
Studienwerk and Deutsche Forschungsgemeinschaft.
\vglue 0.5cm
{\elevenbf\noindent 6. References \hfil}
\vglue 0.4cm


\begin{thebibliography}{99}
\bibitem{FT} For a review see S. Ferrara and S. Theisen,
 {\elevenit Moduli Spaces, Effective Actions and Duality Symmetry in String
 Compactifications}, preprint CERN--TH 5652/90 (1990).
\bibitem{DKL} L. J. Dixon, V. Kaplunovsky and J. Louis,
              {\elevenit Nucl.\ Phys.} {\elevenbf B355} (1991) 649; \\
              J. P. Derendinger, S. Ferrara, C. Kounnas and F. Zwirner,
              {\elevenit Nucl.\ Phys.} {\elevenbf B372} (1992) 145; \\
              P. Mayr and S. Stieberger,
              {\elevenit Threshold Corrections to Gauge Couplings in
              Or\-bi\-fold Compactifications},
              Munich preprint MPI--PH/93--07 (1993).
\bibitem{IKNQ} L. E. Ib\'a\~nez, J. E. Kim, H. P. Nilles and F. Quevedo,
               {\elevenit Phys.\ Lett.} {\elevenbf 191B} (1987) 282.
\bibitem{DHVW} L. Dixon, J. A. Harvey, C. Vafa and E. Witten,
               {\elevenit Nucl.\ Phys.} {\elevenbf B261} (1985) 678 and
                                       {\elevenbf B274} (1986) 285; \\
               L. E. Ib\'a\~nez, J. Mas, H. P. Nilles and F. Quevedo,
               {\elevenit Nucl.\ Phys.} {\elevenbf B301} (1988) 157.
\bibitem{EJN} J. Erler, D. Jungnickel and H. P. Nilles,
               {\elevenit Phys.\ Lett.} {\elevenbf 276B} (1992) 303.
\bibitem{EK} Jens Erler and Albrecht Klemm,
             {\elevenit Comment on the Generation Number in Orbifold
                          Compactifications},
             Munich preprint MPI--Ph/92--60 and TUM--TH--146/92 (1992)
             to appear in {\elevenit Comm.\ Math.\ Phys.}
\bibitem{MOP} D. G. Markushevich, M. A. Olshanetskii and A. M. Perelomov,
              {\elevenit Comm.\ Math.\ Phys.} {\elevenbf 111} (1987) 247.
\bibitem{MS1} M. Spali\~nski, {\elevenit Nucl.\ Phys.} {\elevenbf B377}
              (1992) 339.
\bibitem{LLW} W. Lerche, D. L\"ust and N. P. Warner,
              {\elevenit Phys.\ Lett.} {\elevenbf 231B} (1989) 417.
\bibitem{LMN} J. Lauer, J. Mas and H. P. Nilles,
               {\elevenit Phys.\ Lett.} {\elevenbf 226B} (1989) 251 and
               {\elevenit Nucl.\ Phys.} {\elevenbf 351B} (1991) 353.
\bibitem{JE} J. Erler, {\elevenit Investigation of Moduli Spaces in String
             Theories}, Thesis at the Technical University of Munich
             (in German language), Munich preprint MPI--Ph/92--21 (1992).
\bibitem{CLO} M. Cvetic, J. Louis and B. Ovrut,
              {\elevenit Phys.\ Lett.} {\elevenbf 206B} (1988) 227.
\bibitem{FFS} S. Ferrara, P. Fr\`e and P. Soriani,
              {\elevenit Class.\ Quant.\ Grav.} {\elevenbf 9} (1992) 1649.
\bibitem{ES} J. Erler and M. Spali\~nski,
             {\elevenit Modular Groups for Twisted Narain Models},
             Munich preprint MPI--Ph/92--61 and TUM--TH--147/92 (1992).
\bibitem{MS2} M. Spali\~nski, {\elevenit Phys.\ Lett.} {\elevenbf 275B}
              (1992) 47.
\bibitem{couplings} S. Hamidi and C. Vafa,
                    {\elevenit Nucl.\ Phys.} {\elevenbf B279} (1987) 465; \\
                    L. J. Dixon, D. Friedan, E. Martinec and S. Shenker,
                    {\elevenit Nucl.\ Phys.} {\elevenbf B282} (1987) 13; \\
                    J. A. Casas, F. Gomez and C. Munoz,
                    {\elevenit Phys.\ Lett.} {\elevenbf 251B} (1990) 99; \\
                    T. T. Burwick, R. K. Kaiser and H. F. M\"uller,
                    {\elevenit Nucl.\ Phys.} {\elevenbf B355} (1991) 689; \\
                    J. Erler, D. Jungnickel, J. Lauer and J. Mas,
                    {\elevenit Ann.\ Phys.}  {\elevenbf 217} (1992) 318; \\
                    J. Erler, D. Jungnickel and J. Lauer,
                    {\elevenit Phys.\ Rev.} {\elevenbf D45} (1992) 3651; \\
                    S. Stieberger, D. Jungnickel, J. Lauer and M. Spali\~nski,
                    {\elevenit Mod.\ Phys.\ Lett.} {\elevenbf A7} (1992)
                    3059; \\
                    J. A. Casas, F. Gomez and C. Munoz,
                    {\elevenit Int.\ J.\ Mod.\ Phys.} {\elevenbf A8} (1993)
                    455;\\
                    J. Erler, D. Jungnickel, M. Spali\~nski and S. Stieberger,
                    {\elevenit Higher Twisted Sector Couplings of Z(N)
                    Orbifolds}, Munich preprint MPI--Ph/92--56 and
                    TUM--TH--142/92 (1992); \\
                    D. Jungnickel, {\elevenit Correlation Functions of
                    Two-dimensional Twisted Conformal Field Theories},
                    Thesis at the Technical University of Munich (in German
                    language), Munich preprint MPI--Ph/92--27 (1992); \\
                    S. Stieberger, {\elevenit Higher Twisted Sector Couplings
                    and Duality Transformations in Orbifold
                    Compactifications}, Dilpoma thesis at the Technical
                    University of Munich (in German language),
                    Munich preprint TUM--TH--148/92 (1992).
\bibitem{COGP} P. Candelas, X. C. De la Ossa, P. S. Green and L. Parkes,
               {\elevenit Nucl.\ Phys.} {\elevenbf B359} (1991) 21 and
               {\elevenit Phys.\ Lett.} {\elevenbf 258B} (1991) 118.
\end{thebibliography}
\end{document}